Saturn's Titan: A Strict Test for Life's Cosmic Ubiquity

Jonathan I Lunine
Lunar and Planetary Laboratory
The University of Arizona
Tucson AZ 85721 USA

Is life a common outcome of physical and chemical processes in the universe? Within our own solar system, a successful search for even primitive life, were it to have an origin independent from life on Earth, would dramatically advance a positive answer (1). The most stringent test for a second independent origin of life would come from examination of either the most physically remote from Earth, or the most exotic type, of planetary environments in which one might plausibly imagine a form of life could exist. In this paper I argue that Saturn's moon Titan is the best such target in our solar system. Further, Titan might be a type example of a planetary environment abundant throughout the cosmos.

The truncated Drake equation

The Copernican revolution of the $15^{th}$ through $17^{th}$ centuries moved the Earth squarely away from the center of the universe and fostered a cosmology in which every aspect of the Earth and solar system was, or was predicted to be, a common facet of the cosmos (2). This "Copernican" assumption about the Universe is confirmed in every step upward in spatial scale, from planetary systems to galaxies to clusters of galaxies to large-scale cosmic structures. Where it seems to fail at present is in comprehending the nature of the Universe relative to other putative "universes"; modern cosmology has difficulty explaining the particular values of fundamental constants and what appears to be a particularly improbable set of starting conditions for the cosmos (3). Thus, while the spirit of Copernicanism was extended to biology in the $19^{th}$ century through the gradually developing understanding that cellular processes work on the same physical and chemical principles as nonliving matter, it could not be applied automatically to the union of the biological and astronomical sciences: that is, to the "astrobiological" question of whether life is common in the cosmos (4).

The problem of our own unique versus common cosmic status as an intelligent and technological life form is symbolized in the Drake equation, first formulated almost 50 years ago (5). The equation expresses the number of observable extraterrestrial civilizations in our Milky Way Galaxy as equal to $R \times f_p \times n_e \times f_l \times f_i \times f_c \times L$, where R is the rate of formation of suitable stars (it is sufficient to assume those similar in mass and composition to the Sun) in our galaxy, $f_p$ the fraction of stars with planets, $n_e$ the average number of such planetary systems with a habitable, or life-sustaining, environment, $f_l$ the fraction of habitable planets on which life actually forms, $f_i$ the fraction of those life-bearing planets with intelligent life, $f_c$ the fraction of those intelligence-bearing planets with a civilization technically capable of transmitting signals, and L the average lifetime of such a civilization. From the point of view of "astrobiological Copernicanism"— whether life itself is a common cosmic phenomenon—it is enough to consider a "truncated Drake equation" of the first four terms only, multiplied by the duration of time over which a planet remains habitable. This is the number of life-bearing planets in the Galaxy.

Much of 20$^{th}$ century astronomy not concerned with cosmology was devoted to establishing the first term in the Drake equation, while the other terms remained speculative. In 1996 it became possible to survey widely for, and detect, planets around other stars. It is now known that approximately 10% or more of stars that are similar to the Sun in their "spectral class" (chemical composition and mass) have planets, though up until 2009 planets the size of Earth could not be detected (6). As of this writing, NASA's Kepler mission has begun a survey that will fully address the second, and partly address the third, term of the Drake equation by surveying for planets the size of our own world, and slightly smaller, around stars of the same spectral class as the Sun. Once completed, should the statistical occurrence of Earth-sized "exoplanets" be high enough, the next step would be to build spaceborne devices sufficiently sensitive to directly study any Earth-sized planets that might be present around the nearest stars to our solar system, detect their atmospheres, and measure their masses (7). Such facilities, however, ambitious, are potentially feasible in the coming two decades and would more fully quantify the third term in the Drake equation.

What remains unquantified in the truncated Drake equation, then, is the frequency of habitable planets that actually host life—or did at one time. Although it is often tacitly assumed that if a planet is habitable it hosts life, this is a Copernican *assumption*. Because the origin of life on Earth is not yet understood in terms of a specific set of chemical steps that led from abiotic organic chemistry to biochemistry, one cannot claim to know the robustness of such a process given variations in planetary environmental conditions—or even under no variation whatsoever from conditions on the Earth before life began. And we will not get the answer from studying Earth-sized planets around other stars, when that becomes feasible: the kinds of spectral signatures available from exoplanets will be either ambiguous in telling us that life exists, or too subtle for even spaceborne systems that plausibly might be built in the coming decades. For example, molecular oxygen can be a signature of photosynthesis (biotic) or of the rapid escape of water from the upper reaches of a planetary atmosphere (abiotic); the simultaneous presence of methane ($CH_4$) makes more likely the biological explanation (8) but methane could be limited in abundance and hence difficult or impossible to see on a planet light years from Earth.

It is remarkable that modern astronomy has quantified or is about to quantify the first three terms in the Drake equation written down a half century ago. But to address the fourth term definitively will be beyond the reach of astronomical observations for the foreseeable future. Two alternative strategies are (a) to determine in the laboratory or computer how life might have begun on Earth and assess the robustness of the chain of physical and chemical processes responsible or (b) find evidence in our own solar system of past or present life that had a separate origin from that of life on Earth. In the rest of this paper I will argue the pros and cons of particular solar system bodies as targets for (b).

Searching for life elsewhere in the solar system

All known forms of life on Earth are consistent with an origin from a so-called last universal common ancestor—a community of cells that shared a common physical history of descent from non-living organic matter (9). Were evidence to be found for a form of life that convincingly required a second, separate origin from life on Earth, it would provide a strong case for the ubiquity of life in the cosmos. The search for such a second form of life in the accessible cosmos is a search for life on a small subset of accessible and plausible targets within our solar system:

*Earth*: One cannot rule out the possibility of a terrestrial biosphere that had a separate origin but is coexistent with the known biosphere. Such a "shadow biosphere"

(10) would provide a convincing second origin argument if it were sufficiently biochemically distinct from us (for example, using right-handed amino acids, or left-handed sugars, or both). Such a search does not require space vehicles but might require accessing remote places in the Earth's crust where "normal" life cannot go. The absence of evidence for such a shadow biosphere might imply the extinction of such life forms at the hands of "normal life" sometime in the past, or that the origin of life on Earth was truly singular.

*Mars:* Relatively accessible from the Earth, Mars' surface environment is likely not habitable because of a strongly oxidizing near-surface layer, lack of persistent surface water and strong exposure to ultraviolet radiation from the Sun. The subsurface crust is a plausibly habitable environment, however, and the recent discovery of methane emission from a few regions on Mars raises the possibility of a subsurface biota (11). Whether the methane emission is from biological or geological processes might be assessed by sufficiently sensitive determination of the ratio of the two stable isotopes of carbon in the methane: biological processes favor incorporation of the lighter isotope in biologically-produced organic molecules, although this test can be ambiguous. Even were this to be possible (by an orbiter or a lander), and biological production of methane confirmed, it would fail to establish that Martian life had a separate origin from that of the Earth: exchange of crustal material between Earth and Mars by hypervelocity impact has been demonstrated through the existence of Martian meteorites on the Earth (12), and such transfer may be survivable by microorganisms (13). Thus, absent direct sampling of (likely deep-seated) subsurface Martian organisms, the discovery of biological processes on the Red Planet does not definitively answer the question at hand.

*Europa:* A favorite target for astrobiologists, this lunar-sized moon of Jupiter possesses an ice crust beneath which is a salty layer of liquid water perhaps 100 km thick (14), or about twice the volume of the Earth's oceans (15). The ocean is maintained by heating associated with the tides raised by Jupiter, was suspected based on the surface geology (16), and indirectly detected through the effect of the salty water on Jupiter's magnetic field (17), Whether life exists within this subsurface ocean is a major question in the field of astrobiology, and NASA has announced that its next major mission to the outer solar system will be to orbit Europa and determine the thickness of the ice crust. That thickness will determine the strategies on a later mission for sampling ocean material and—if present—life.

A positive detection of life on Europa would not by itself address the question of life's *cosmic* ubiquity. Most astrobiologists assign high priority in the search for life to locations that could be inhabited by organisms from our own planet, because these are the only form of life we know. It is therefore essential to assess whether signs of life in an alien environment are the result of an origin independent from Earth, or are contamination from Earthly life that colonized such an environment—be it yesterday or 4 billion years ago. Indeed, for this reason the Committee on Space Research (COSPAR) of the International Council of Scientific Unions has a strict set of guidelines for the sterilization of planetary spacecraft to avoid contaminating potentially habitable environments with terrestrial organisms, although sterilization standards for Martian landers have varied from one nation to another. Europa is deemed by COSPAR to deserve the highest level of planetary protection (19); NASA's Galileo Jupiter orbiter—not sterilized—was deliberately directed into a destructive entry into Jupiter's atmosphere in 2003 when its fuel supplies were nearly depleted so as to avoid a future contaminating collision with Europa.

Life from Earth (or Mars, for that matter) could arrive at Europa via hypervelocity impacts; terrestrial bacterial spores are known to last sufficiently long to survive such transfer (18) and the subsequent impact (19). The greater distance of Europa relative to Mars, and the high velocities of impact thanks to Jupiter's gravity, make survival of hitchhiking organisms less likely but not impossible (20).

The biggest challenge to detecting life on Europa is accessibility. It is generally assumed now that direct sampling would be of organisms preserved in the ice crust rather than in the ocean itself, which would be inaccessible to available or plausible technologies even were it only hundreds of meters thick, well below the geophysical estimates which range from a few to tens of kilometers thickness (21). However, this assumes that organisms would be preserved in a recoverable fashion, and that the ice layers where such organisms exist would be cycled near enough to the surface to be sampled without being so close that the prodigious particle radiation would destroy the organic remains. Even so, delivery and operation of a vehicle designed for the surface of Europa, given the intense particle radiation flux from the Jovian magnetosphere, would be a challenge.

*Enceladus:* In some ways potentially a miniature Europa, this small icy moon of Saturn emits active plumes of material containing water ice, ammonia and a host of organic molecules measured by the Cassini Saturn orbiter (22). Enceladus is also the major source of Saturn's E-ring; some of the ice particles within that ring contain sodium (23), suggesting that its source region within Enceladus contains liquid water today or did in the recent past. If indeed liquid water persists in some part of the subsurface of Enceladus, its presence along with organic molecules, nitrogen and mineral salts provides a promising abode for terrestrial-type life. Here again, one cannot completely rule out the possibility of cross-contamination with the Earth, albeit a small one. Enceladus is easier to explore than Europa, with an external radiation environment vastly more quiescent, and a much smaller gravitational field. However, the location of any liquid water pockets, and their persistence with time, remain as questions that would need to be resolved prior to attempting a mission designed to penetrate the surface and test for life in Enceladus' interior.

*Titan*: The second-largest moon in the solar system, Titan has a nitrogen-rich atmosphere four times denser at its surface than Earth's (24), a wealth of organic molecules in the atmosphere and on the surface (dominated in the atmosphere by methane) (25), and a "hydrologic cycle" in which methane is the main working fluid (26). The joint US-European mission called "Cassini-Huygens" has, since 2004, mapped Titan's surface with imagers, radar, the Huygens descent probe, and other instruments, revealing the presence of equatorial dunes made of organic particles, channels carved in what is probably a water-ice landscape, and high latitude lakes of methane and ethane (27). This exotic landscape—water ice geology carved through the action of liquid methane and ethane—is the result of Titan's large size and great distance from the Sun, stabilizing the atmosphere and ensuring a very cold (-179º Celsius) environment suitable for liquid methane and ethane.

Such conditions, in which water is completely frozen-out and present as liquid only for short times after meteorite impacts and geologic heating events, would seem unfavorable for life. And, indeed, for life as it is known to exist on Earth—life that requires liquid water—only a possible liquid water layer deep beneath Titan's ice crust would seem habitable (28). Such a layer is unreachable given the present-day crustal thickness at or exceeding 50 km by most estimates, although clues to the presence of life might come from chemical and isotopic anomalies (29). Alternatively, one might ask whether the terrestrial biological requirement for liquid water is too restrictive. Could the liquids that are stable at Titan's surface—methane and ethane—play host to a form of organic chemistry that would include all the attributes we associated with life? Such a form of life would be so different from terrestrial life that one would be forced to conclude it had an independent origin.

<u>The possibility of life in hydrocarbon lakes on Titan</u>

The hundreds of lakes seen on Titan by the Cassini orbiter instruments cover approximately 15% of Titan's known surface above $65^{o}$ north (only about half of this north polar region has been imaged), and less in the southern hemisphere (30). However, this is a substantial physical area, given that Titan is larger than the Earth's moon, and the three largest lakes are referred to as "mare" by the international Astronomical Union in recognition of their size (figure 1). While the evidence for liquids in the lakes and seas is circumstantial, it is derived from several distinct types of observations (31). The seas may be tens of meters deep based on their very dark appearance to the radar system onboard Cassini (32).

The ambient conditions on Titan are such that the dominant constituents of the lakes and seas are likely to be liquid ethane and methane, with a small admixture of nitrogen from the atmosphere (roughly 1%), and lesser amounts of other organic molecules that rain down continuously into the lakes from the upper atmosphere where solar ultraviolet rays and charged particles from the Saturn environment act on the methane. Thus, Titan's lakes and seas are quite different from those of the Earth, but they are bodies of liquid nonetheless, and the largest should be stable for very long periods of time. There is even the possibility that the seas are connected to an even larger subterranean body of liquid methane, ethane, or both (30).

Could such lakes and seas, containing hydrocarbon liquid rather than water, host something that might be considered life? Much of the work to address this question has been done by S. Benner and his colleagues (33), and the rest of this section is a summary of their conclusions. Because all life on Earth requires liquid water, and indeed the fundamental polymers of biology—proteins and the nucleic acids—require intimate contact with liquid water for their proper functioning, it is assumed that liquid water has properties that make it especially suited for life. Indeed, water is liquid over a broad range of temperatures, and water ice floats on liquid water providing an insulating effect when temperatures hover around freezing. Water forms bonds between its oxygen atom and a hydrogen atom in a neighboring water molecule ("hydrogen bond") more readily than it does with most organic molecules. This allows the formation in water of membranes composed of organic molecules, as well as proper folding of proteins. When one observes the interaction between known biopolymers and water, the impression is that water is uniquely suited as a solvent for biology.

This impression is, however, based on an intrinsic (and powerful) bias. The biology that works in liquid water is the biology that we see—because water is the dominant liquid stable under terrestrial conditions. But Benner and colleagues have pointed out some undesirable properties of water such as, for example, the insolubility of the reactive form of carbon dioxide and the elaborate ways life has evolved to deal with it (33).

Might then liquid hydrocarbons, stable under Titan conditions, be a suitable biosolvent? Methane and ethane have their own problems, including their non-polarity, which means that as liquids they provide no support for molecular structures that depend on interaction with the liquid for their stability. But small amounts of polar molecules might exist in the Titan seas. Furthermore, an interesting "bio"chemistry might be built around the dominance of hydrogen bonding between organic molecules immersed in the non-hydrogen-bonding ethane and methane (S. Benner, pers. comm.), and in such a biochemistry, the low temperatures and consequent slow reaction rates are not necessarily a disadvantage (J. Nott, pers. comm.). While nothing like a complete, theoretical biochemistry in liquid methane and ethane has been constructed, there is no particular property or set of properties of liquid methane and ethane that could lead one to a priori rule out in such a medium a kind of self-sustaining, replicating, catalytic organic chemistry that might be called life.

Titan as a test for the ubiquity of life in the cosmos

I have deliberately not tackled the question of what constitutes life, and indeed whether it is possible to make such a definition, (34) for want of space. This makes it hard to generalize from what are regarded as the absolute requirements for life on Earth to the requirements for any chemical system we might deem worthy of being called alive.

At a minimum, if we adopt organic chemistry as the basis for life, the requirements might be a fluid environment, a source of free energy and abundant organic molecules—all of which Titan possesses (35).  But it is a great leap to go from terrestrial biology, with its water-based organic chemistry at room temperature, to the cold, hydrocarbon-soaked poles of Titan.

And that is precisely my point. Were one to sample the hydrocarbon lakes and seas of Titan and find structures or patterns in the organics present there that suggest chemical cycles which generate their own catalysts (figure 2), or information-carrying molecules that reliably generate the same catalysts and structures over and over again, one might conclude that a second form of life had been found. And because it was found in a solvent medium completely alien—inhospitable—to terrestrial life, one could confidently conclude that it had a separate origin from life on Earth. Although Titan, like Europa, receives debris from hypervelocity impacts on Earth (20), any contaminating terrestrial life form, introduced by such impacts, would not survive the cryogenic hydrocarbon lakes of Titan. This eliminates planetary-protection concerns that plague plans for the exploration of Europa and Mars. Furthermore, life based on covalent bonds (us) and life based on hydrogen-bonding are so chemically different that the former cannot be the ancestor of the latter.  Thus, even the most primitive kind of self-organizing organic chemistry in the lakes of Titan—an early step on the road to life—would be a momentous find. It would tell us that life began independently multiple times in our solar system, and so the fourth term in the truncated Drake equation could well be closer to unity than to zero.

Titan as a model for the most common "habitable planet"

The conclusion of the previous section was admittedly provocative—but only because we know of only one kind of life, that which exists in liquid water on a planet orbiting 1 astronomical unit (150 million miles) from a so-called G-dwarf star, our Sun. Since we do not know how life formed on the Earth, we cannot extrapolate what we know of terrestrial biochemistry to other environments. But we can explore other environments, and in the next section I will argue that Titan is particularly easy to explore. But if Titan is a special case, is it really worth exploring? Does it really inform us about the value of the fourth term ($f_l$) in the Drake equation if the third term—the fraction of planets that are habitable—pertains only to planets like the Earth around stars like the Sun?

The most common type of stable star in the cosmos is not the G-dwarf—a star like the Sun—but the much less massive "M", or red dwarf. M dwarfs are long-lived and, because they range from 1/10 to ½ the mass of the Sun, are much less luminous than our home star. A typical M dwarf is so much less bright than the Sun (36) that a planet with stable liquid water on its surface would have to orbit roughly 0.1 AU from the brightest M dwarfs—ten times closer than the Earth is to the Sun. Many large planets have been found around other stars at such close orbital distances (indeed they are easier to detect with current techniques), and so it would seem there should be no problem with imagining a plenitude of Earth-like habitable worlds in tight orbits around the universe's most common type of star.

However, the close proximity (0.1 AU and inward) of a planet to an M dwarf raises a host of problems. Such planets are tidally locked to their parent star, presenting only one face to the source of free energy needed for life, are exposed to intense flares and stellar winds leading to loss of atmosphere (37), and may lack an amount of water sufficient to sustain a hydrosphere by virtue of having formed so close to the star (38). Whether these effects rule out life is unknown, but they certainly create complications in trying to use a simple definition of orbital distance to define a habitable, Earth-like environment—one on whose surface liquid water is stable—for M dwarfs. An Earth-sized body, even with adequate water, orbiting at 0.1 AU from an M dwarf will *not* have an environment resembling that on our home world.

Conditions at 1 AU—the Earth-Sun distance—from an M dwarf are much less severe and hence more favorable for a stable environment on a planet. A planet at 1 AU from an M dwarf can rotate decoupled from its orbital motion, as does the Earth, evening out the flux of light from the parent star. Flares and stellar winds at 1 AU from an M dwarf would present no more threat to a putative planet than we face from the comparable phenomena emitted from the Sun.

Because M dwarfs are so much less bright than the Sun, the same physics that dictates a world with stable liquid water must be only 0.1 AU from an M dwarf parent says that at 1 AU from such a star should be a planet like Titan: at that distance the light from the red dwarf is so faint as to maintain liquid methane and ethane on a planet's surface, but not liquid water. And because M dwarfs outnumber G-dwarfs by a factor of between ten and a hundred, one can conclude that this situation—a Titan-like world at 1 AU—may be far and away the more common cosmic situation than our own.[1]

---

[1] The fact that Titan is a moon of Saturn is largely immaterial to the nature of its surface environment. Were Titan orbiting the Sun at Saturn's distance, it would possess essentially the same surface and atmospheric properties, being little affected by Saturn itself. However, a planet formed around an M dwarf at 1 AU would not necessarily have a mass close to that of Titan—it could have the mass of the

Therefore the search for an exotic, even if very primitive, form of life in the hydrocarbon lakes and seas of Titan has potentially profound implications for the cosmic ubiquity of life. A positive answer would force the third term in the Drake equation—$n_e$—traditionally defined in terms of the environment of our own Earth, to be radically expanded to include the kind of environment present on Titan.

### Future exploration of the Titan environment

Titan's great distance from the Earth is its only disadvantage when it comes to exploration. Everything else about Titan: its low gravity, dense atmosphere, calm low altitude winds, low radiation levels—are distinct advantages over other solar system targets of interest for life. A robotic probe deployed by the European Space Agency successfully entered the atmosphere, descended, and landed on Titan's surface in 2005, returning data from the air and the surface (39). Replicating this mission in the high northern latitudes, in one of the great seas of Titan, with a probe instrumented to detect life, is eminently feasible. To cover large areas of Titan's diverse surface, a more ambitious hot air balloon whose buoyancy is maintained by the waste heat of the vehicle's power source has been carefully studied by US and European space agencies (40), and by private entrepreneurs such as Julian Nott; it would be a kind of floating rover taking full advantage of Titan's dense atmosphere and benign environment (figure 3).

---

Earth, or it could be a giant planet like Saturn or Jupiter. An Earth-mass body at 1 AU from an M dwarf the age of the Sun could have very different surface properties from Titan, because the contribution of geothermal energy to surface heating will be more significant than for Titan, though declining progressively with time. And that such a planet in the epoch of its formation could garner sufficient methane from a primordial disk around an M dwarf to create large-scale lakes and seas is only a plausible supposition. I do not mean to minimize such complications, but their proper treatment would require a treatise vastly longer than this work.

Designing experiments to detect life in an alien environment constitutes a rich field of research unto itself (41). It is sufficient here to point out that a series of elemental, chemical and isotopic tests, coupled with an imaging microscope to examine the lake material, would be required to assess whether organic chemistry in the methane-ethane medium is being mediated in a sophisticated way indicative of life (42). At the same time, given the difficulty in searching for terrestrial-type life in exotic environments, the challenge of finding an exotic form of life in an alien environment cannot be minimized.

Conclusion

Titan has been ranked by a minority of astrobiologists as the highest priority place to go search for life in our solar system, both for the potential to find exotic life that inarguably had an independent origin from us, and the relative ease of operation on Titan's surface (1). I have recapitulated these arguments and added what I believe is a novel one, namely that planetary environments somewhat like Titan's may be far more common than Earth-like environments in the cosmos. This argument hinges on three points: (i) M dwarfs are the most common stable (hydrogen-burning) star in the cosmos; (ii) Earth-like temperatures on a planetary surface are achievable only in very close proximity to an M-dwarf, where the tidal and energetic environment would create decidedly non-Earthlike, and possibly unstable, environments; (iii) a Titan-like planetary environment is achieved at orbital distances from an M-dwarf which are comparable to the Earth-Sun distance, and where conditions should be stable.

The dichotomy I have laid out here between Earth-like and Titan-like planets is an obvious oversimplification: one might imagine a diversity of planets around M dwarfs including those that are Europa-like, Mars-like, or for which no solar system analog exists whatsoever. And all potential life-sustaining environments available for study in our solar system should be searched. But of these *only Titan provides both easy accessibility to the potential habitable environment of interest and an assurance that—should life be found—its origin is independent from that of life on Earth.*

Saturn's moon Titan, therefore, provides a potentially strong assessment of the magnitude of the final term in the truncated Drake equation presented at the start of this article. It provides a fishing license to broaden the search for planets around other stars from those that have Earth-like surface conditions to those (at ~1 AU from M dwarfs) that are Titan-like, to assess how common such environments might be. Should the methane-ethane seas of Titan host an exotic form of life, one not based on the same biochemistry as that in water on the Earth, the question of whether life is or is not a common cosmic phenomenon will have been largely answered.

Acknowledgement

I am grateful to chemist Steven Benner and aeronautical explorer Julian Nott for stimulating discussions. This work was supported by the NASA Astrobiology Institute.

Figures

Figure 1: Strips of radar imagery from the Cassini mission showing the three large "mare" (seas) believe to be basins of ethane and methane liquid. Ligeia Mare is about 500 km in diameter and is centered on about 79°N latitude; Kraken Mare, based on these and other less detailed images, may be as large as the Earth's Caspian sea. Gaps in the coverage occur because the radar system on Cassini cannot be steered, but rather covers whatever terrain the spacecraft flies over as it passes close to Titan in its orbital tour around Saturn. NASA/JPL/ASI radar data with names established by the International Astronomical Union overlain.

Figure 2. Abundances of alkanes—a type of hydrocarbon, or carbon-hydrogen organic molecule—are shown for two samples. The number of carbon atoms in each alkane peak is labeled. The top panel shows the end product of a simple abiotic chemical process. The bottom panel is an analysis of a petroleum product, and hence the product of biological processes. The abiotic sample shows a broad, smooth distribution of abundance; the biological sample shows an uneven distribution of alkanes and two in particular, phytane and pristine, which are found to be especially abundant in biological processes. The challenge for Titan would be to identify equivalent signatures of an exotic, non-aqueous biochemistry. Top panel from (43); figure from (44).

Figure 3: (a) A Montgolfière, or hot air balloon, floats high above a methane-ethane lake on Titan. The power source for the instruments provides heat for buoyancy, which in turn is regulated by computer-controlled opening and closing of flaps on the balloon. Painting by Tibor Balint, JPL/Caltech. (b) The European Space Agency Huygens Probe, which landed on Titan in January 2005, is seen embedded within its foil-covered heat shield during testing. A probe of similar size and design could splashdown into one of Titan's hydrocarbon seas, and float indefinitely. European Space Agency.

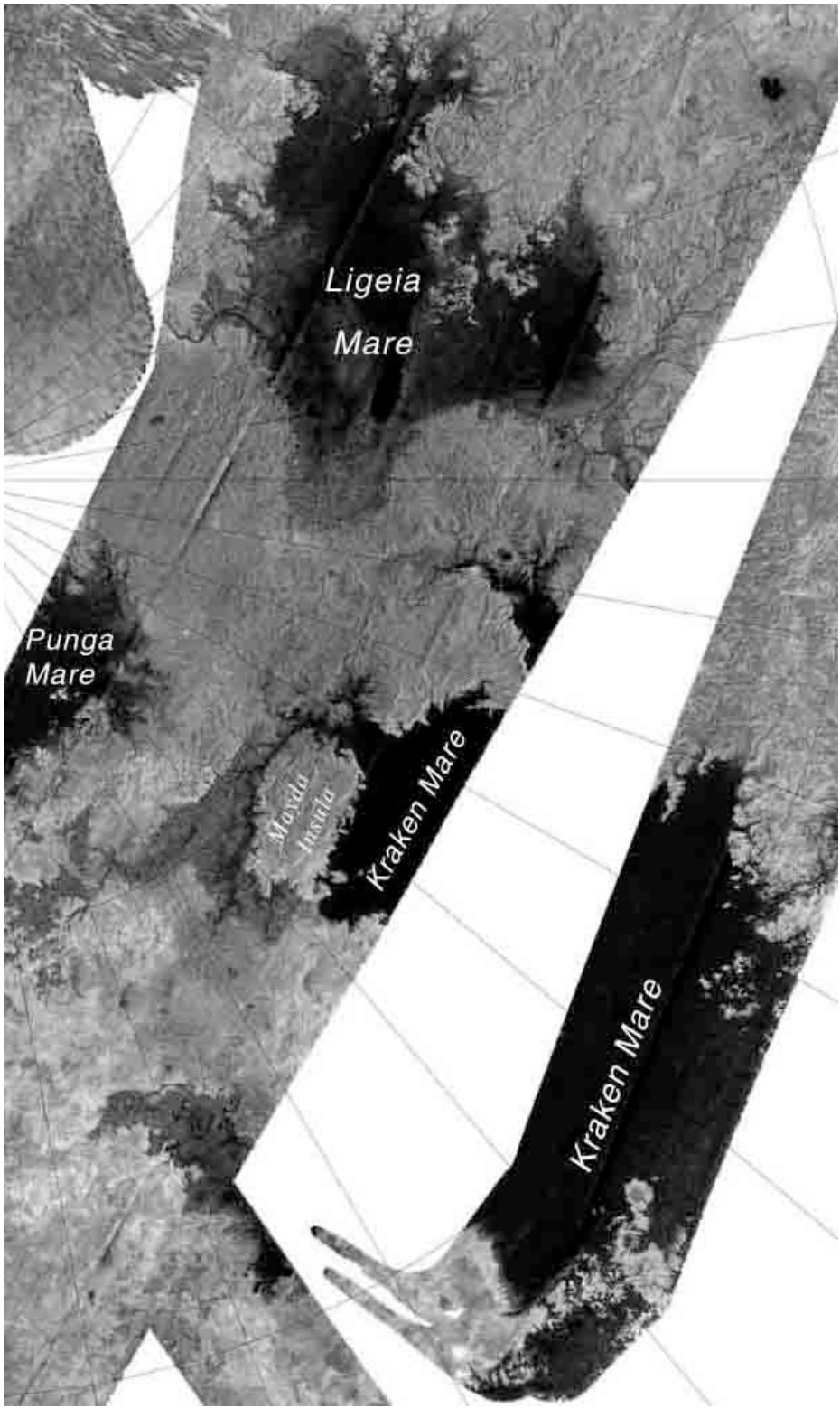

Figure 1

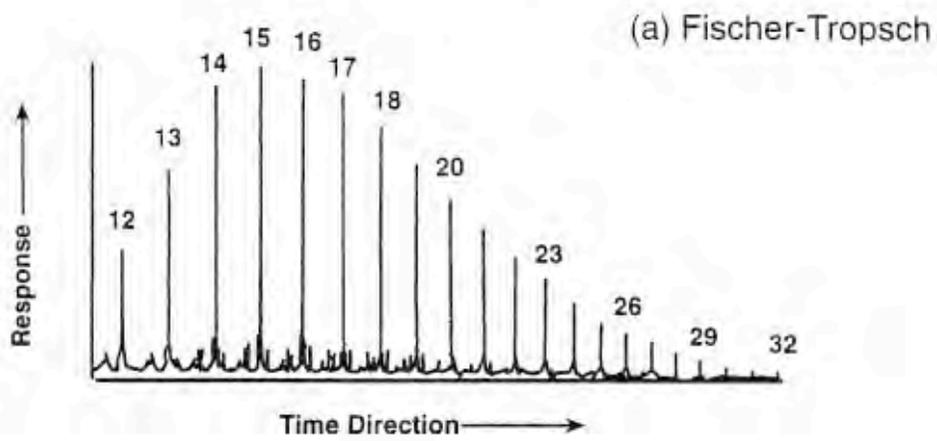

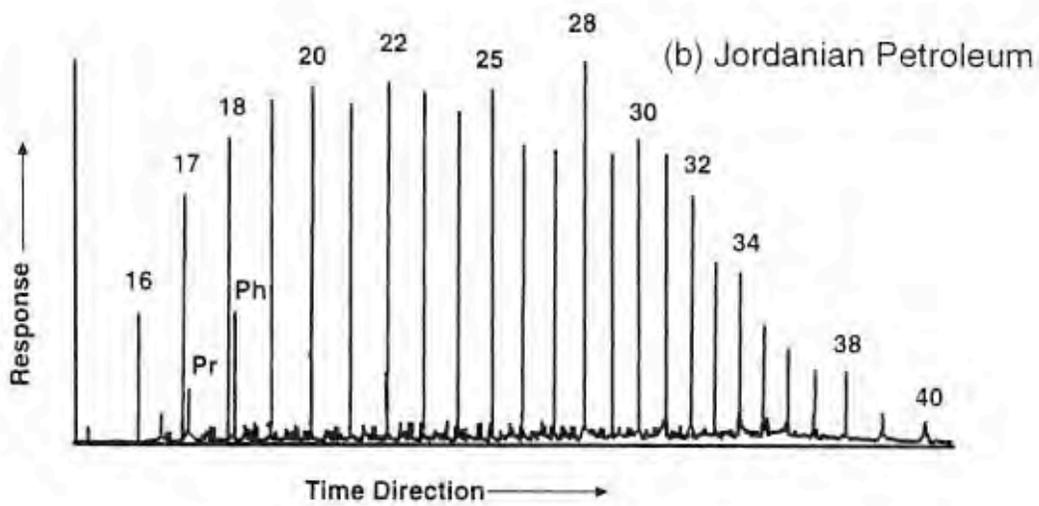

Figure 2

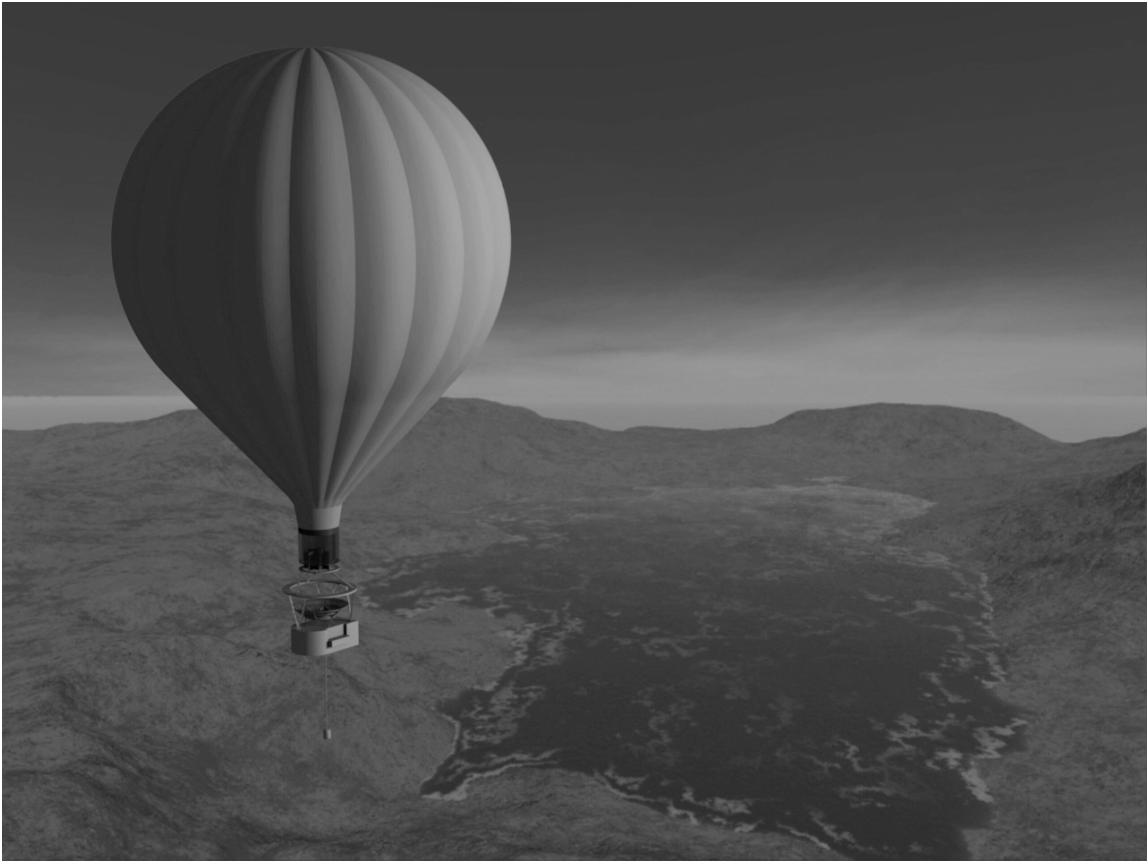

Figure 3(a)

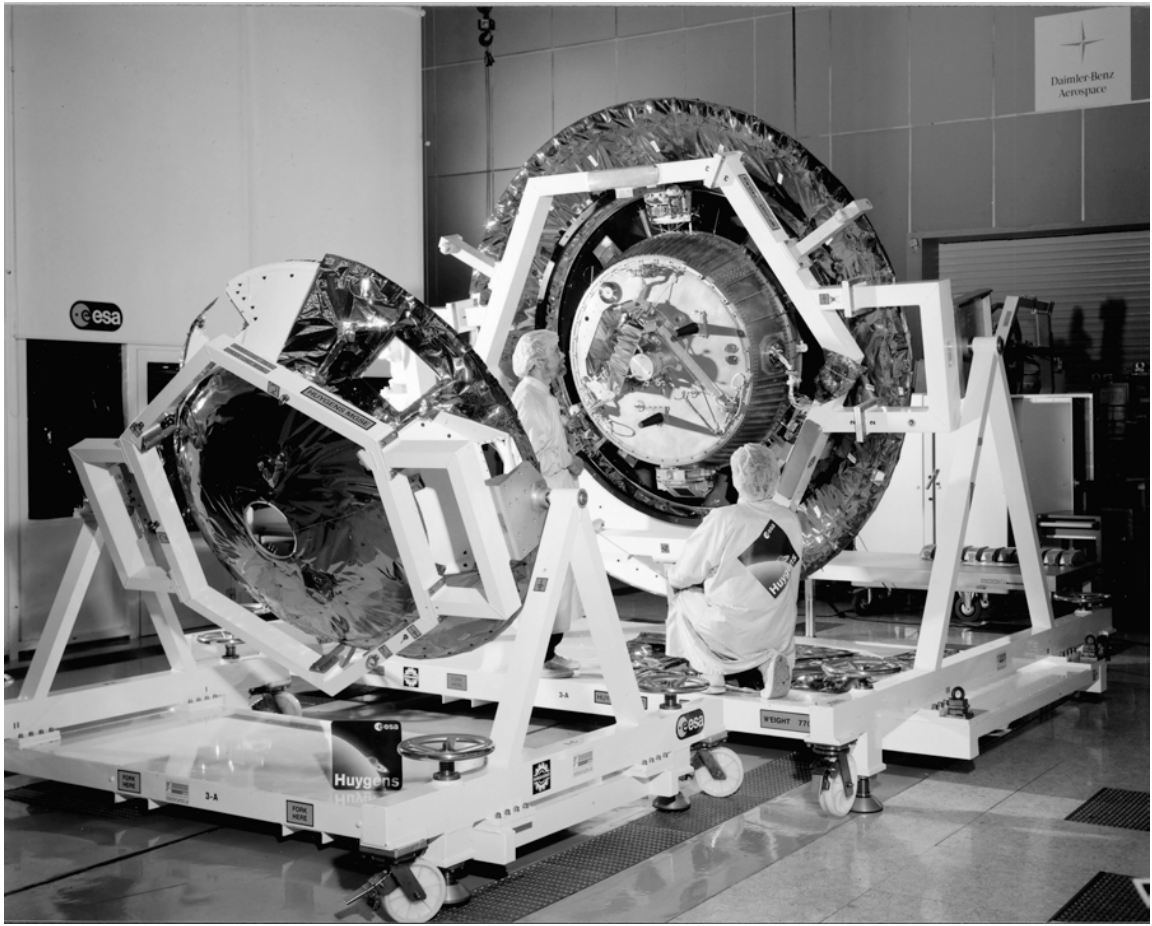

Figure 3(b)